\documentclass[aps,prb,twocolumn,showpacs,superscriptaddress]{revtex4}
\usepackage{epsfig}
\usepackage{graphicx}
\usepackage{amsfonts}
\usepackage[figuresright]{rotating} 
\usepackage{amssymb}
\usepackage{amsmath}

\def\avg#1{\langle#1\rangle}

\def\be{\begin{equation}} \def\ee{\end{equation}}
\def\bea{\begin{eqnarray}} \def\eea{\end{eqnarray}}

\def\nn{\nonumber}

\begin{document}
\title{Unconventional Bose-Einstein condensations from spin-orbit coupling}

\author{Congjun Wu}
\affiliation{Department of Physics, University of California, San Diego,
CA 92093}
\author{Ian Mondragon-Shem}
\affiliation{Department of Physics, University of California, San Diego,
CA 92093}
\affiliation{ Instituto de  F\'{i}sica, Universidad de
Antioquia, AA 1226, Medell\'{i}n, Colombia}
\author{Xiang-Fa Zhou}
\affiliation{\textit{Key Laboratory of Quantum Information,
University of Science and Technology of China,  Hefei, Anhui
230026, China}}

\begin{abstract}
According to the {\it ``no-node''} theorem, many-body ground state 
wavefunctions of conventional Bose-Einstein condensations (BEC) are 
positive-definite, thus time-reversal symmetry cannot be spontaneously broken.
We find that multi-component bosons with spin-orbit coupling provide an 
unconventional type of BECs beyond this paradigm.
We focus on the subtle case of isotropic Rashba spin-orbit coupling and 
the spin-independent interaction.
In the limit of the weak confining potential, the condensate wavefunctions 
are frustrated at the Hartree-Fock level due to the degeneracy of the
Rashba ring.
Quantum zero-point energy selects the spin-spiral type condensate through the 
``order-from-disorder'' mechanism.
In a strong harmonic confining trap, the condensate spontaneously generates
a half-quantum vortex combined with the skyrmion type of spin texture.
In both cases, time-reversal symmetry is spontaneously broken.
These phenomena can be realized in both cold atom systems with artificial
spin-orbit couplings generated from atom-laser interactions and 
exciton condensates in semi-conductor systems.
\end{abstract}
\maketitle

The conventional many-body ground state wavefunctions of bosons satisfy the 
celebrated {\it ``no-node''} theorem in the absence of rotation, as written
in Feynman's textbook\cite{feynman1972}, which means that they are 
positive-definite in the coordinate representation.
This theorem implies that time-reversal (TR) symmetry cannot be 
spontaneously broken.
It applies to various ground states of bosons
including Bose-Einstein condensates (BEC), Mott-insulating states,
density-wave states, and supersolid states, thus making this a very
general statement. 

It would be exciting to search for novel types of quantum ground
states of bosons beyond the ``no-node'' paradigm, such as unconventional
BECs with complex-valued wavefunctions.
This theorem does not apply to spinful bosons with spin-orbit 
(SO) coupling, whose linear dependence on momentum
invalidates Feynman's proof.
Artificial SO coupling from laser-atom interactions 
have been applied to ultra-cold boson systems
and its effects have been investigated. 
\cite{lin2009,lin2009a,juzeliunas2008,
vaishnav2008,stanescu2007,stanescu2008,spielman2009,lin2011}.
On the other hand, excitons in semiconductors 
\cite{snoke1990,butov2004a,butov2007,timofeev2007} 
exhibit SO coupling in their center-of-mass motion
\cite{hakioglu2007,can2008,yao2008}.
In particular,  exciting progress has been made in indirect exciton 
systems in coupled quantum wells where electrons and holes are
spatially separate \cite{butov1994,butov2002}.
The extraordinarily long life-time of indirect excitons provides 
a wonderful opportunity to investigate 
exciton condensation \cite{butov2004a} with SO coupling.

In this article, we show that spin-orbit coupled bosons
develop unconventional BECs beyond the ``no-node'' theorem.
The Rashba SO coupled BEC with the spin-independent
interaction exhibits frustration at the Hartree-Fock level.
Quantum zero-point fluctuations select a coherent condensation
in the presence of weak spatial inhomogeneities, which exhibits
spiral spin-density waves and spontaneous TR symmetry breaking.
In a strong external harmonic trap, the ground state condensate 
develops orbital angular momentum, which can be viewed as a 
half-quantum vortex. 
Moreover, the spin density distribution exhibits a cylindrically
symmetric spiral pattern as skyrmions.

We begin with a 3D two-component boson system with Rashba SO 
coupling in the $xy$-plane and with the contact spin-independent
interaction, described by
\bea
H&=&\int d^3 \vec r~~ \psi^\dagger_\alpha
\Big\{ -\frac{\hbar^2 \nabla^2}{2M}- \mu \Big\} 
\psi_\alpha +\hbar\lambda_R \psi^\dagger_\alpha \Big\{
-i \nabla_y \sigma_x  \nn \\
&+& i \nabla_x \sigma_y\Big\} \psi_\beta
+\frac{g}{2}\psi^\dagger_\alpha 
\psi^\dagger_\beta \psi_\beta \psi_\alpha, ~~~~~
\label{eq:mnbdy}
\eea
where $\psi_\alpha$ is the boson operator; the pseudospin indices 
$\alpha=\uparrow, \downarrow$ refer to two different internal components 
of bosons; $\lambda_R$ is the SO coupling strength which carries 
the unit of velocity;
$g$ describes  the $s$-wave scattering interaction.
Eq. (\ref{eq:mnbdy}) possesses a Kramer-type TR symmetry $T=i\sigma_2 C$ 
satisfying $T^2=-1$,  where $C$ is the complex conjugate operation and 
$\sigma_2$ operates on the boson pseudospin degree of freedom. 

In the homogeneous system, the single particle
states are the helicity eigenstates of $\vec \sigma \cdot 
(\vec k \times \hat z)$ with a dispersion relation given by
$\epsilon_{\pm} (\vec k)= \frac{\hbar^2}{2 M} [(k \mp k_{so})^2+k_z^2]$
where $k_{so}=\frac{M\lambda_R}{\hbar}$.
The energy minima are located on the lower branch along a ring
with the radius $k_{so}$ in the plane of $k_z=0$. 
The corresponding two-component wavefunction $\psi_{+}(\vec k)$ 
with $|k|=k_{so}$ can be solved as
$\psi_{+} (\vec k)= \frac{1}{\sqrt 2} (e^{-i \phi_k/2}, i e^{i \phi_k/2})^T$,
where $\phi_k$ is the azimuth angle of the projection of $\vec k$
in the $xy$-plane.
The interaction part in Eq. (\ref{eq:mnbdy}) in the helicity basis 
can be represented as
\bea
H_{int}&=&\frac{g}{2}\sum_{\lambda\mu\nu\rho}\sum_{p_1p_2q}
\avg{\vec p_1+\vec q; \lambda| \vec p_1; \rho}
\avg{\vec p_2-\vec q; \mu | \vec p_2; \nu} \nn \\
&\times&\psi^\dagger_\lambda (\vec p_1+\vec q) 
\psi^\dagger_\mu (\vec p_2-\vec q)
\psi_\nu (\vec p_2) \psi_\rho (\vec p_1),
\eea
where the Greek indices $\lambda, \nu, \mu, \rho$ are the helicity 
indices $\pm$; the matrix elements denote the inner product of 
spin wavefunctions of two helicity eigenstates at different momenta,
e.g., $\avg{\vec p_1+\vec q; \lambda| \vec p_1; \rho}=
\frac{1}{2} [1 +\lambda\rho e^{i(\phi_{p_1}-\phi_{p_1+q})} ]$.

The low energy Rashba ring brings degeneracy for the 
condensate configurations, {\it i.e.}, frustrations.
Bosons tend to avoid the positive exchange energy for repulsive 
interactions, which is the driving force for BECs.
Spin polarizations at two opposite ends of a diameter of
the low energy ring are orthogonal to each other, and thus condensations
with these states are free of  exchange interactions.
Without loss of generality, we define the fragmented
and coherent condensates of $\Phi_{frag}$ and $\Phi_{coh}$, respectively,
as
\bea
\Phi_{frag} &=&\frac{1}{\sqrt{N_{A} ! N_{B} !}  }
[ \psi^\dagger_+(\vec k_A) ]^{N_{A}}
[ \psi^\dagger_+(\vec k_B) ]^{N_{B}}
|0\rangle, \\ 
\Phi_{coh}&=& \frac{1}{\sqrt {N_0!}}\Big\{ \sqrt {\frac{n_A}{n_0}}
\psi^\dagger_+(\vec k_A)+ e^{i\phi}\sqrt{\frac{n_B}{n_0}}
\psi^\dagger_+(\vec k_B)\Big \}^{N_0} |0\rangle \nn \\
\label{eq:coh}
\eea
where $A$ and $B$ are points with
$\vec k_A=(-k_{so},0,0)$ and $\vec k_B=(k_{so},0,0)$; 
$(N_{A}, N_{B})$ is the particle number partition 
satisfying $N_{A}+N_{B}=N_0$ with $N_0$ the total particle 
number in the condensate; $n_{A,B}=N_{A,B}/V$, and 
$n_0=N_0/V$; the phase $\phi$ in Eq. \ref{eq:coh} can be 
absorbed by the shift of the origin.
At the Hartree-Fock level, $\Phi_{frag}$ and $\Phi_{coh}$ with an 
arbitrary partition $N_{A,B}$ have the same energy.
Since fragmented condensates with different
$N_{A,B}$ carry different momenta, they do not mix in the ideal
homogeneous systems.
However, even very weak spatial inhomogeneity can build up coherence
among them, and leads to coherent condensates \cite{mueller2006,ho2000}. 
Below we consider coherent condensates $\Phi_{coh}$ in Eq. \ref{eq:coh},
and leave a detailed study of the competition 
between fragmented and coherent condensates to a later work.

\begin{figure}
\centering\epsfig{file=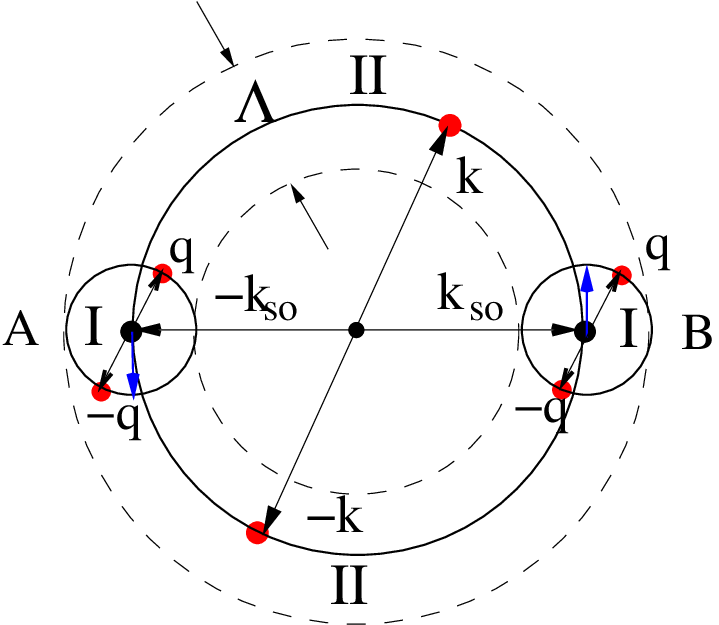,clip=1,width=0.75\linewidth,angle=0}
\caption{The low energy ring with the radius $k_{so}$ in momentum space.
The coherent condensate involves points $A$ and $B$ with orthogonal 
spin polarizations.
The low energy Bogoliubov excitations 
within $|k-k_{so}|<\Lambda$, $|k_z|<\Lambda$, 
and $\Lambda/k_{so}\ll 1$ are classified into two regimes 
I (inside two cylinders centering around $A$ and $B$ with radius
of $\Lambda$) and II (outside). 
}
\label{fig:ring}
\end{figure}

The zero-point quantum fluctuations lift the degeneracy among
coherence condensates with different partitions of $n_{A,B}$.
We define a momentum scale $k_{int}$ satisfying
$\hbar^2 k_{int}^2/(2M)=gn_0$ where $gn_0$ is the interaction energy 
scale.
We only consider the limit of strong SO coupling, {\it i.e.}, 
$k_{in}\ll k_{so}$, and leave the general case for a later study.
We chose an intermediate momentum cutoff $\Lambda$ satisfying 
$k_{int}\ll \Lambda \ll k_{so}$, and study the Bogoliubov excitations
in a cylindrical shell of
\bea
|\sqrt {k_x^2+k_y^2}-k_{so}|<\Lambda, ~~ |k_z|<\Lambda.
\label{eq:cutoff}
\eea
Within this shell, interaction energy is stronger than the
kinetic energy, thus particle and hole states are significantly
mixed.
We further divide this shell into two parts I and II as depicted 
in Fig. \ref{fig:ring}.
Part I is inside two cylinders with the radius of $\Lambda$
centering around points $A$ and $B$, and part II is outside 
these two cylinders.

For part I, we define boson operators in the lower branch as 
$a_{\vec q}=\psi_+(-k_{so} \hat e_x+\vec q)$ and $
b_{\vec q}=\psi_+(k_{so} \hat e_x+\vec q)$, and $\avg{a_{q=0}}
=\sqrt{N_{a}}$ and $\avg{b_{q=0}}=\sqrt{N_{b}}$, respectively.
The low energy excitations in this region have been calculated in 
Ref. [\onlinecite{stanescu2008}].
By defining 
$\gamma^\dagger_1(\vec q)=\frac{1}{\sqrt {N_0}}(\sqrt{N_a} a^\dagger_q 
+\sqrt{N_b} b^\dagger_q )$, 
$\gamma^\dagger_2(\vec q)=\frac{1}{\sqrt {N_0}}(\sqrt{N_b} a^\dagger_q 
-\sqrt{N_a} b^\dagger_q )$, 
the mean-field Hamiltonian, up to the order of $q^2$, is represented as
\bea
H_{MF,1}&=&\sum_q \Big\{ E(\vec q) \gamma^\dagger_1(\vec q) \gamma(\vec q)
+  gn_0 [\gamma^\dagger_1 (\vec q) \gamma^\dagger_1 (-\vec q) +h.c.]
\nn \\
&+&E(\vec q) \gamma_2^\dagger(\vec q) \gamma_2 (\vec q) \Big \},
\eea
where $E(\pm \vec q)\approx \hbar (q_x^2+q_z^2)/(2M)$ up to
the order of $O(q^3/k_{so})$.
The Bogoliubov modes mixing $\gamma_1^\dagger(\vec q)$ and $\gamma_1(-\vec q)$
exhibit the spectrum of
$\hbar\omega(\vec q)=\sqrt{E_{\vec q} (E_{\vec q} + 2 g n_0)}
\approx \sqrt{\frac{\hbar g n_0}{M}} \sqrt{q_x^2+q_z^2}$.
This is the phonon mode describing the overall
density fluctuations, which exhibits linear dispersion
relation for $\vec q$ in the $xz$-plane and becomes soft for 
$\vec q$ along $\hat e_y$.
The $\gamma_2$ mode represents the relative density fluctuations between two
condensates which describes spin wave excitations.
This mode remains a free particle spectrum $E(\vec q)$. Both
of the $\gamma_{1,2}$ modes only depend on the total 
condensation fraction $N_0$. Hence, the contribution
from part I does not lift the degeneracy between different
partitions of $(N_A,N_B)$ up to the quadratic order of $q$.

Next we turn to the Bogoliubov spectra in part II where 
$\psi_+(\vec k)$ is degenerate with  $\psi_-(-\vec k)$
but not with $\psi_+(2 \vec k_{A,B} -\vec k)$.
The mean-field Hamiltonian reads
\bea
H_{MF,2}&=&\sum_k \psi^\dagger_+(\vec k) \psi_+(\vec k) 
\big\{ \epsilon(k) +\frac{g}{2} (n_0-\Delta n \cos \phi_k)\big\} \nn \\
&+& g \sqrt{n_a n_b}\big\{\psi^\dagger_+(\vec k) \psi^\dagger_-(-\vec k)
\cos\phi_{\vec k} e^{i\phi_{\vec k}}+h.c. \big\}, ~~~~
\eea
where  $\Delta n =n_a-n_b$.
The Bogoliubov spectra can be solved as $H_{MF,2}=\sum_{\vec k}
\big\{\omega (\vec k) (\gamma_3^\dagger (\vec k) \gamma_3 (\vec k) 
+\frac{1}{2}) \big\}$ with
\bea
\omega(k)&=&\sqrt{\epsilon_{\vec k} (\epsilon_{\vec k}+ g n_0)
+\frac{g^2n_0^2}{4} 
f(x)} + \frac{g n_0}{2} x \cos\phi_k, ~~~
\label{eq:dispersion}
\eea
where $x=\Delta n/n_0$ and $f(x)= \sin^2\phi_k + x^2 \cos^2\phi_k$.
The second term in Eq. \ref{eq:dispersion} averages to zero,
and thus the total zero-point energy in regime II 
depends on $x^2$.
It reaches the minimum at $x=0$, or, $n_a=n_b$,
which describes a spin-density-wave 
spiral in the $xz$-plane with the condensate wavefunction as
\bea
\psi_{cond}=(\cos k_{so} x, \sin k_{so} x)^T.
\label{eq:cond}
\eea
An accurate evaluation of the zero-point energy needs to
deal with the ultraviolet divergence of the integral over
momentum space, which will be postponed to a later publication.

The above ``order-from-disorder'' results can be captured from constructing
an effective Gross-Pitaevskii (GP) equation.
Generally speaking, the interaction parameters in the GP equation
are renormalized from their bare values in Eq. \ref{eq:mnbdy}
by the zero-point motions.
Because the kinetic energy only possesses the $SO(2)$ rotational symmetry,
although the bare interaction is spin-independent,
an extra spin-dependent term should be generated as
\bea
g^\prime \Big[n_\uparrow(\vec r)- n_\downarrow(\vec r)\Big]^2,
\label{eq:extra}
\eea
where $n_{\uparrow,\downarrow}$ are particle densities of two 
spin components.
Obviously, $g^\prime<0$ selects the spin-spiral condensate
involving the $s_z$ component (e.g. Eq. \ref{eq:cond} ) 
even though the total $s_z$ averages to zero,
while $g^\prime>0$ selects the plane-wave condensate 
with spin polarization in the $xy$-plane (e.g. 
$\psi_{cond}= \frac{1}{\sqrt 2}e^{ik_{so} x} (1,i)^T$).
According to the result of Eq. \ref{eq:cond}, we conclude that
$g^\prime<0$ is the case for  Eq. \ref{eq:mnbdy}.
Furthermore, $g^\prime$ is at the order of $g^2$ from the power-counting
of the integral of the zero-point energy.
The usual leading order correction to the interaction parameter
in 3D BECs is of $g^\frac{5}{2}$.
This difference is due to the effective dimension
reduction by the Rashba ring. 

Next we consider the situation of a strong confining potential 
$V_{ex}(r)=\frac{1}{2} M \omega_T^2 (x^2+y^2)$ and a relatively 
weak interaction.
The condensate along the $z$-axis is set uniform.
In this case, the single particle energy dominates over the
interaction energy, which  mixes the plane-wave states
along the low energy ring.
We define the SO energy scale $E_{so}= \hbar \lambda_R /l$ where 
$l=\sqrt{\hbar/(M\omega_T)}$ is the length scale of the trap,
and the dimensionless parameter $\alpha=E_{so}/(\hbar \omega_T) =l k_{so}$. 
Let us gain some intuition in the strong SO limit $\alpha\gg 1$.
The harmonic potential in the momentum representation
becomes  $V_{ex}=\frac{1}{2} 
M \omega_T^2 (i\partial_{\vec k}-\vec A(\vec k))^2$, 
where $\vec A(\vec k)=i\avg{\psi_+(\vec k)|
\partial_{\vec k} |\psi_+(\vec k)}$ carrying a 
$\pi$-flux located at $\vec k =(0,0)$.
$V_{ex}$ quantizes the orbital motion around the ring as
\bea
\Delta E_{m+\frac{1}{2}}= \frac{1}{2} M \omega_T^2 
(\frac{m+\frac{1}{2}}{k_{so}})^2 = \frac{1}{2 \alpha^2} \hbar \omega_T
(m+\frac{1}{2})^2.
\eea
The single particle ground state forms the Kramer doublets corresponding to 
$m+\frac{1}{2}=\pm\frac{1}{2}$.
Equivalently, in the real space,
due to the 2D rotational symmetry, the  wavefunctions
can be denoted by the total angular momentum $j_z=m+\frac{1}{2}$. 
The ground state single particle wavefunctions form a Kramer doublet 
as represented in cylindrical coordinates as 
\bea
\psi_{\frac{1}{2}}=\left( \begin{array}{c}
f(r) \\
g(r) e^{i\phi}
\end{array}
\right), ~~~
\psi_{-\frac{1}{2}}=\left( \begin{array}{c}
-g(r) e^{-i\phi} \\
f(r)
\end{array}
\right), 
\label{eq:TR_doublet}
\eea
\noindent 
where $f(r)$ and $g(r)$ are real functions.
At $\alpha\gg 1$, these doublet states have nearly equal weight in the spin 
up and down components, {\it i.e.}, $\int dr d\phi~ r|f(r)|^2 \approx 
\int dr d\phi~ r |g(r)|^2$,  thus the spin moment averages to zero. 
In the presence of weak interactions,
bosons condense into one of the TR doublets, the average orbital
angular momentum per particle is $\hbar/2$, i.e., one spin component 
stays in the $s$-state and the other one in the $p$-state.
This is a half-quantum vortex configuration 
spontaneously breaking TR symmetry \cite{salomaa1985, wu2005, zhou2003}.

\begin{figure}
\centering\epsfig{file=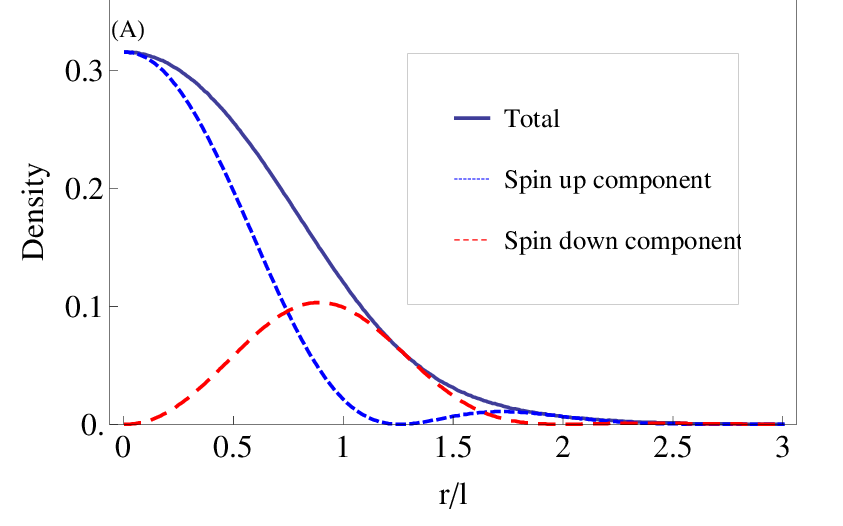,clip=1,width=0.9\linewidth,
 angle=0}
\centering\epsfig{file=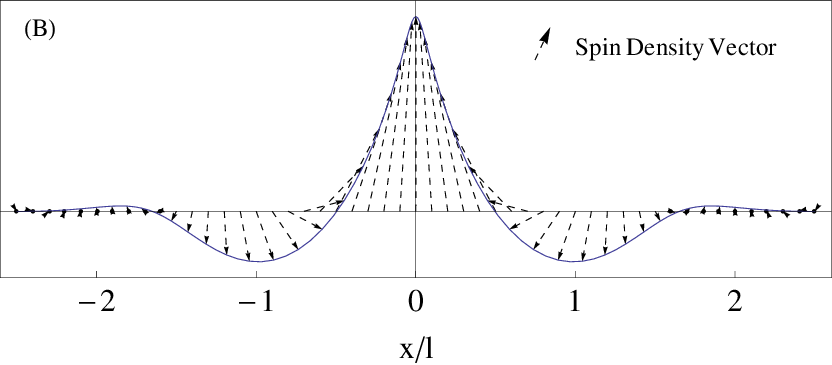,clip=1,width=0.9\linewidth,
height=0.3\linewidth,angle=0}
\caption{(A) The radial density distribution of spin up and down
components, and the total density distribution in the unit of $N_0$ 
at $\alpha=2$ and $\beta=5$. 
(B) The skyrmion type spin texture configuration plotted in the
$xz$-plane. 
}
\label{fig:trap}
\end{figure}

We have performed numerical calculation to confirm the above
intuitive picture.
The characteristic interaction energy scale is defined as $E_{int}=
g N_0/(\pi l^2 L_z)$, where $L_z$ is the system size 
along the $z$-axis, and the dimensionless parameter
$\beta=E_{int}/(\hbar \omega_T)$.
We numerically solve the GP equation 
\bea
\Big\{
-\frac{\hbar^2  \nabla^2}{2M} &+&
\hbar \lambda_R(
-i \nabla_y \sigma_{x,\alpha\beta}  
+ i \nabla_x \sigma_{y,\alpha\beta})
+g n(r,\phi) \nn\\
&+&\frac{1}{2}M \omega^2_T r^2\Big\} \psi_\beta(r,\phi)
=E \psi_\alpha(r,\phi),
\label{eq:GPtrap}
\eea
where $n(r,\phi)$ is the particle density.
The parameter values are chosen as $\alpha=2$ and $\beta=5$.
The interaction effectively weakens the harmonic potential and 
does not change the orbital partial wave structure of the wavefunctions.
We show the radial density profiles of both spin components 
$|f(r)|^2$ and $|g(r)|^2$ in Fig. \ref{fig:trap} A.
Each of them exhibits oscillations at a pitch value of approximately
$ 2 k_{so}$, which originate from the low energy ring structure and, thus, 
are analogous to the Friedel oscillations in fermion systems.
The spin density,  defined as $\vec S (r,\phi)= 
\psi_{\alpha}^* (r,\phi) \vec 
\sigma_{\alpha\beta} \psi_{\beta}(r,\phi)$, exhibits an
interesting topological spin texture configuration.
Let us first look at its distribution along the $x$-axis where
the supercurrent is along the $y$-direction and the spin lies in
the $xz$-plane.
Explicitly, we express $S_z(r,\phi)=\frac{1}{2}
(|f(r)|^2-|g(r)|^2)$ and $S_x(r,\phi)=f(r)g(r)$.
The radial oscillations of $|f(r)|^2$ and $|g(r)|^2$ have an approximate
$\pi$ phase shift, which arises from the different angular
symmetries.
As a result, $\vec S$ spirals in the $zx$-plane along the $x$-axis as
plotted in Fig. \ref{fig:trap} B 
at the pitch value of the density oscillations.
The spin density distribution in the whole space can be obtained through a 
rotation around the $z$-axis, which exhibits the skyrmion
configuration.

Because of the non-linearity of the GP equation, the superposition principle, 
generally speaking, does not apply.
Nevertheless, if we only keep the $SU(2)$ invariant 
interaction term in the GP equation, all of the linear superpositions 
of the Kramer doublet $\psi_{\pm\frac{1}{2}}$ in Eq. \ref{eq:TR_doublet} 
can rotate into one another, thus are degenerate.
Therefore, in real experiment systems, if the initial state is prepared 
with total angular momentum $j_z=0$, we will obtain 
a superposition of $\psi_{\pm\frac{1}{2}}$ due to the conservation of $j_z$.
On the other hand, if the initial state is prepared with 
the average $j_z$ per particle $\pm\frac{1}{2}$, say, by cooling down
from the fully polarized spin up or down state, then $\psi_{\pm\frac{1}{2}}$ will 
be reached. 
We also remind that if we go beyond the Hartree-Fock level
to include the zero-point motion correction, 
the extra spin-dependent term of Eq. \ref{eq:extra}
will also lift the above accidental degeneracy.

\begin{figure}
\centering\epsfig{file=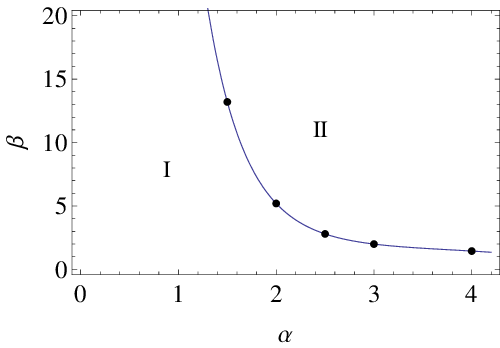,clip=1,width=0.8\linewidth,
 height=0.5\linewidth}
\caption{The phase diagram boundary of $\beta_c$ v.s $\alpha$ between 
(I) the half-quantum vortex condensate and (II) spin-spiral condensate.
The transition from (I) to (II) breaks rotational symmetry.
}
\label{fig:transition}
\end{figure}

So far we have presented two different types of condensations.
The half-quantum vortex condensate preserves rotational symmetry, which 
is favored by the trapping potential and is stable for weak 
interaction strengths.
Instead, the spin-spiral condensate involving two plane-waves with 
opposite wavevectors breaks rotational symmetry.
It is favored by interactions, and should survive under weak
trapping potentials. 
Or, equivalently, with fixing $\alpha$, it can be stabilized
by increasing the interaction energy scale $\beta$.
We have performed the numerical study for the critical
line between them as presented in Fig. \ref{fig:transition}.
For simplicity, the spin-independent interaction is still used.
We calculate the expectation value of $\avg{G|j^2_z|G}$ of the 
condensate wavefunction.
In regime I, the condensate is chosen as the eigenstate with
$j_z=\frac{1}{2} (-\frac{1}{2})$ and thus $\avg{G|j^2_z|G}=\frac{1}{4}$.
In regime II, $\avg{G|j^2_z|G}$ deviates from $\frac{1}{4}$.
The condensate starts to involve high angular momentum components, and thus 
breaks rotational symmetry.
It is qualitatively in the same phase of spin-spiral 
condensate with cylindrical boundary condition.

The recent research focus of the ``synthetic gauge fields'' in cold atom 
systems provides a promising method to observe the above exotic SO coupled 
BECs \cite{lin2009,lin2009a,spielman2009,lin2011,campbell2011}.
In particular, anisotropic SO coupled BEC has been realized 
experimentally \cite{lin2011}.
Furthermore, several proposals of realizing isotropic Rashba 
SO coupling have appeared \cite{campbell2011,stanescu2007,zhang2011}.
A typical set of parameter values are provided in Ref. \cite{stanescu2007}
for a trap frequency $\omega_T=2\pi \times 10$ Hz,
a characteristic Rabi frequency $\Omega= 2\pi \times 10^7$ Hz,
and a detuning $\Delta=2\pi \times 10^{11}$Hz,
which satisfy $\omega_T\ll \Omega \ll \Delta$ and $\omega_T\ll \
\Omega^2/\Delta$. 
The value of $\alpha$ can vary from $0$ up to the order of several tens.
With the typical particle number $N_0=10^6$, the interaction energy scale
$E_{int}$ is around $100nK$, or $2$ kHz \cite{leggett2001}, and
$\beta$ is of the order of several tens.
Thus for the most convenient experimental parameters, the spin-spiral 
condensate is realized.
Nevertheless, the half-quantum vortex condensate can be realized 
by reducing the condensate particle number by one order to $N_0=10^5$
combined with increasing the trap frequency.
The detection would be straightforward as performed in 
previous time-of-flight imaging in vortex experiments 
\cite{madison2000,anderson2000}.
By separately imaging the density profiles of the pseudospin
up and down components in the time-of-flight spectra
\cite{madison2000}, the radial oscillation of the $S_z$
component can be directly seen.

Another class of SO coupled boson systems are the indirect 
excitons in 2D coupled double quantum wells.
Both electrons and holes possess relativistic SO coupling and 
so does the center of mass motion of excitons.
The real space spin configurations of exciton condensations can be 
conveniently detected through photoluminescence from electro-hole 
recombination.
Excitingly, the recent experiment has observed spin-textures 
of the coherent exciton systems which arise from SO coupling
and exhibit a similar pattern shown in Fig. \ref{fig:trap}
\cite{high2011}.
We leave a detailed study on this phenomenon to a future paper. 

In summary, we find that bosons with SO coupling exhibit 
complex-valued condensations beyond Feynman's ``no-node'' paradigm.
The coherent spin-spiral BEC is realized when
interaction energy is dominant, while the half-quantum 
vortex BEC is stable when the trapping potential is strong.
The half-quantum vortex condensate also exhibits
the skyrmion type topological spin-texture configuration.

C. W. is supported by U.S. NSF-DMR1105945 and AFOSR-YIP program.
X. F. Z. acknowledges the support of CUSF, SRFDP
(20103402120031), and the China Postdoctoral Science.
C. W. thanks helpful discussions with L. Butov, L. M. Duan, M. Fogler,
J. Hirsch, T. L. Ho,  L. Sham, S. C. Zhang, and F. Zhou.

{\it Note added:}
After the three version of this manuscript was posted on arXiv,
there appeared two experimental works of SO coupled BECs
in both cold atom and exciton systems \cite{lin2011,high2011},
and several theoretical investigations
\cite{ho2010,wang2010,yip2010,zhang2011,xu2011,kawakami2011}.


\begin{thebibliography}{10}

\bibitem{feynman1972}
R.~P. Feynman, {\em Statistical Mechanics, A Set of Lectures} (Addison-Wesley
  Publishing Company, ADDRESS, 1972).









\bibitem{juzeliunas2008}
G. Juzeliunas {\it et~al.}, Phys. Rev. Lett. {\bf 100},  200405  (2008).

\bibitem{vaishnav2008}
J.~Y. Vaishnav and C.~W. Clark, Phys. Rev. Lett. {\bf 100},  153002  (2008).

\bibitem{stanescu2007}
T.~D. Stanescu, C. Zhang, and V. Galitski, Phys. Rev. Lett. {\bf 99},
  110403  (2007).

\bibitem{lin2009}
Y.-J. Lin {\it et~al.}, Phys. Rev. Lett. {\bf 102},  130401  (2009).

\bibitem{lin2009a}
Y.-J. Lin {\it et~al.}, Nature {\bf 462},  628  (2009).

\bibitem{spielman2009}
I.~B. Spielman, Phys. Rev. A {\bf 79},  063613  (2009).


\bibitem{lin2011}
Y. J. Lin, K. Jimenez-Garcia, and I. B. Spielman,
Nature {\bf 471}, 83 (2011).


\bibitem{stanescu2008}
T. Stanescu, B. Anderson, and V. Galitski, Phys. Rev. A {\bf 78},  023616
  (2008).

\bibitem{snoke1990}
D.~W. Snoke, J.~P. Wolfe, and A. Mysyrowicz, Phys. Rev. B {\bf 41},  11171
  (1990).

\bibitem{butov2004a}
L.~V. Butov, J. Phys.: Cond. Matt. {\bf 16},  R1577  (2004).

\bibitem{butov2007}
L.~V. Butov, J. Phys.: Cond. Matt. {\bf 19},  295202  (2007).

\bibitem{timofeev2007}
L.~A.~V. Timofeev {\it et. al}, J. Phys.: Cond. Matt. {\bf 19},
  295209  (2007).

\bibitem{hakioglu2007}
T. Hakioglu and M. Sahin, Phys. Rev. Lett. {\bf 98},  166405  (2007).

\bibitem{can2008}
M.~A. Can and T. Hakioglu,  arXiv:0808.2900, 2008.

\bibitem{yao2008}
W. Yao and Q. Niu, 
Phys. Rev. Lett. 101, 106401 (2008) 

\bibitem{butov1994}
L.~V. Butov {\it et~al.}, Phys. Rev. Lett. {\bf 73},  304  (1994).

\bibitem{butov2002}
L.~V. Butov, A.~C. Gossard, and D.~S. Chemla, Nature {\bf 418},  751  (2002).

\bibitem{mueller2006}
E.~J. Mueller, T.~L. Ho, M. Ueda, and G. Baym, Phys. Rev. A {\bf 74},  33612
  (2006).

\bibitem{ho2000}
T.~L. Ho and S.~K. Yip, Phys. Rev. Lett. {\bf 84},  4031  (2000).

\bibitem{salomaa1985}
M.~M. Salomaa and G.~E. Volovik, Phys. Rev. Lett. {\bf 55},  1184  (1985).

\bibitem{wu2005}
C. Wu, J.~P. Hu, and S.~C. Zhang, Int. J. Mod. Phys. B V24, 311 (2010).

\bibitem{zhou2003}
F. Zhou, Int. Jour. Mod. Phys.B, 17 {\bf 17},  2643  (2003).

\bibitem{larson2009}
J. Larson and E. Sj\"oqvist, Phys. Rev. A {\bf 79},  043627  (2009).


\bibitem{campbell2011}
D. L. Campbell,  G. Juzeliunas, and I. B. Spielman,
arXiv:1102.3945.

\bibitem{leggett2001}
A.~J. Leggett, Rev. Mod. Phys. {\bf 73},  307  (2001).

\bibitem{madison2000}
K.~W. Madison, F. Chevy, W. Wohlleben, and J. Dalibard, Phys. Rev. Lett. {\bf
  84},  806  (2000).

\bibitem{anderson2000}
B.~P. Anderson, P.~C. Haljan, C.~E. Wieman, and E.~A. Cornell, Phys. Rev. Lett.
  {\bf 85},  2857  (2000).

\bibitem{maialle1993}
M.~Z. Maialle, E.~A. de~Andrada~e Silva, and L.~J. Sham, Phys. Rev. B {\bf 47},
   15776  (1993).

\bibitem{high2011}
A. A. High, {\it et al.}, arXiv:1103.0321.

\bibitem{ho2010}
T. L. Ho, S. Z. Zhang, arXiv:1007.0650.

\bibitem{wang2010}
C. Wang, C. Gao, C. M. Jian, H. Zhai, 
Phys. Rev. Lett., 105, 160403 (2010).

\bibitem{yip2010}
S. K. Yip, arXiv:1008.2263.

\bibitem{zhang2011}
Y. Zhang, L. Mao, C. W. Zhang, arXiv:1102.4045.

\bibitem{xu2011}
Z. F. Xu, R. Lv, L. You, Phys. Rev. A 83, 053602 (2011).

\bibitem{kawakami2011}
T. Kawakami, T. Mizushima, K. Machida, arXiv:1104.4179.

\end{thebibliography}

\end{document}